\documentclass[aps,twocolumn,superscriptaddress,floatfix,nofootinbib]{revtex4-1}
\usepackage{graphicx,amsmath,amssymb,verbatim,color}
\usepackage{booktabs}
\usepackage{comment}
\usepackage{soul}
\usepackage[dvipsnames]{xcolor}
\usepackage{bm}
\usepackage[utf8]{inputenc}
\usepackage[colorlinks=true,citecolor=blue,linkcolor=blue,urlcolor=blue, backref=false,pdfborder={0 0 0}]{hyperref}
\usepackage{float}
\usepackage{multirow}
\newcommand{\be}{\begin{equation}\begin{gathered}}
\newcommand{\ee}{\end{gathered}\end{equation}} 
\newcommand{\barr}{\begin{eqnarray}}
\newcommand{\earr}{\end{eqnarray}} 
\usepackage{romannum}
\usepackage[normalem]{ulem}

\begin{document}

\pagenumbering{arabic}

\title{Efficient analytic approximation for small-scale non-cold relic perturbations}
\author{Nanoom Lee}
\email{nanoom.lee@jhu.edu}
\affiliation{William H. Miller III Department of Physics \& Astronomy, Johns Hopkins University, Baltimore, Maryland 21218, USA}
\author{Yacine Ali-Ha\"imoud}
\email{yah2@nyu.edu}
\affiliation{Center for Cosmology and Particle Physics, Department of Physics, New York University, New York, New York 10003, USA}
\author{Marc Kamionkowski}
\email{kamion@jhu.edu}
\affiliation{William H. Miller III Department of Physics \& Astronomy, Johns Hopkins University, Baltimore, Maryland 21218, USA}

\begin{abstract}
We develop a highly accurate analytic approximation for small-scale non-cold relic perturbations by solving the collisionless Boltzmann equation in the quasi-stationary regime. The approximation is implemented in \texttt{CLASSIER} (\texttt{CLASS} Integral Equation Revision), a modified version of the Boltzmann solver \texttt{CLASS} that replaces the traditional truncated Boltzmann hierarchy of non-cold relic multipoles with a small set of integral equations solved iteratively. Applying it to massive neutrinos yields a factor-of-two reduction in total runtime relative to \texttt{CLASSIER} without the approximation. Compared to standard \texttt{CLASS} runs (with $\ell_{\rm max}^{\rm NCDM}=40$ and no late-time massive neutrino fluid approximation) under the same precision setting, \texttt{CLASSIER} with this approximation is faster by a factor of 3--6. The approximation faithfully reproduces the late-time behavior of massive neutrino perturbations and preserves sub-$0.1\%$ accuracy in the matter power spectrum today up to comoving wavenumber $k=100\,{\rm Mpc}^{-1}$. With this approximation, massive-neutrino perturbations are no longer the computational bottleneck on small scales for linear-theory predictions. The approach can be readily extendable to non-standard dark-matter models, and offers prospects for further efficiency gains in high-precision cosmological analyses.
\end{abstract}

\maketitle

\section{Introduction}

Cosmological observations across multiple probes from cosmic microwave background (CMB) anisotropies to the clustering of galaxies are increasingly sensitive to the properties of relic particles beyond standard cold dark matter. Non-cold dark matter (NCDM) or non-cold relics, such as massive neutrinos or other light thermal relics, free-stream on cosmological scales and imprint characteristic signatures on the growth of structure \cite{Lesgourgues:2006nd, Wong:2011ip, Abazajian:2013oma}. The suppression of power on small scales, the scale-dependent growth of matter perturbations, and the induced anisotropic stress in the early Universe all make NCDM a key target for current and next-generation cosmological surveys \cite{LSSTScience:2009jmu, Abazajian:2019eic}.

The standard approach to modeling the evolution of NCDM perturbations is to solve the Boltzmann hierarchy: a formally infinite set of coupled differential equations for the multipole moments of the phase-space perturbation \cite{Ma:1995ey}. In practice, this hierarchy must be truncated at some finite multipole $\ell_{\max}^{\rm NCDM}$, with closure relations imposed at the truncation point. While accurate, this procedure can become computationally expensive at high precision or on small scales, since a large number of multipoles must be evolved. In addition, if the hierarchy is truncated too low, small numerical artifacts may appear, which can propagate into predictions for cosmological observables \cite{class}. To mitigate these challenges, model-specific approximations have also been developed \cite{Hu:1998kj,Lewis:2002nc,Shoji:2010hm,Lesgourgues:2011rh}, most notably for massive neutrinos, which are approximated as a viscous fluid once free-streaming suppresses higher-order multipoles. These approximations substantially reduce computational cost while retaining accuracy on large scales, but they are reliable only for the range of wavenumbers and model parameters that they are calibrated upon, and have a precision no better than the calibration calculations. More recently, improved fluid approximations have been proposed that incorporate a scale-dependent sound speed and anisotropic stress, further extending the regime of validity for massive neutrinos \cite{Nascimento:2023psl}. Nonetheless, high-accuracy calculations of NCDM perturbations still demand evolving a large number of multipoles, which remains a heavy computational expense in modern cosmological analyses. The situation can be even more demanding in certain non-standard dark matter models, where accurately capturing the perturbations requires dense momentum grids in addition to many multipoles, further increasing the computational cost (e.g., see Ref.~\cite{FrancoAbellan:2021sxk}).

Recently, alternative formulations based on integral equations have been developed as a way to overcome these limitations \cite{Kamionkowski:2021njk,Ji:2022iji, Lee:2025zym}. Instead of evolving the entire Boltzmann hierarchy, the integral approach follows particles along unperturbed geodesics and expresses the perturbations as time-integrals over metric sources convolved with spherical Bessel kernels. This method is similar in spirit to the line-of-sight approach for CMB anisotropies \cite{Seljak:1996is}, except that it is used not just to compute final observables, but throughout the evolution of perturbations themselves. An identical integral approach has also been used to follow the evolution of massive neutrinos during non-linear structure formation \cite{2013MNRAS.428.3375A, 2018MNRAS.481.1486B}. This representation naturally captures the physics of free-streaming, and its convolutional structure allows for efficient numerical evaluation using fast Fourier transforms (FFTs). The \texttt{CLASSIER} code (\texttt{CLASS} Integral Equation Revision)\footnote{Available at \href{https://github.com/nanoomlee/CLASSIER}{https://github.com/nanoomlee/CLASSIER}.} \cite{Lee:2025zym} implements this method within the public Boltzmann solver \texttt{CLASS v3.2.2} \cite{class, Lesgourgues:2011rh}, achieving accurate and efficient evolution of NCDM perturbations.

Despite this progress, evaluating the full integral solutions can remain costly at large wavenumbers, where perturbations oscillate rapidly due to free-streaming. In this regime, analytic approximations are both physically transparent and computationally advantageous: they capture the asymptotic behavior of perturbations while reducing the need for expensive numerical convolutions and preserving the accuracy of the observables of interest.

In this work, we develop such a small-scale approximation for NCDM perturbations, in the form of a perturbative expansion in the ratio of the perturbation wavelength to the free-streaming lengthscale. We emphasize that this approximation is a fully analytic, though approximate, solution to the collisionless Boltzmann equation. It is highly accurate yet computationally efficient, and does not rely on any numerical fits or calibrations. We implement this approximation in \texttt{CLASSIER}, apply it to the case of massive neutrinos, and demonstrate that it improves the total runtime of the code by a factor of two, with even larger gains expected as cosmological analyses push to smaller scales. We also validate the accuracy of the approximation against fully numerical solutions, showing that it faithfully reproduces both the oscillatory features and the late-time asymptotics of the multipole moments. Our results highlight the utility of the small-scale limit as a complementary tool to integral-equation methods, providing further efficiency gains for cosmological modeling in the precision era.

The rest of this paper is organized as follows. In Section~\ref{sec:approximation}, we derive the small-scale approximation for non-cold relic perturbations. In Section~\ref{sec:implementation-results}, we describe its implementation in \texttt{CLASSIER} and assess the accuracy and performance of the approximation when applied to massive neutrino, by comparing with fully numerical solutions. We conclude in Section~\ref{sec:conclusion}.

\section{Analytic small-scale approximation for NCDM perturbations}
\label{sec:approximation}

In this Section, we derive analytic approximations to the collisionless Boltzmann equation in the small-scale, free-streaming regime. 

\subsection{Collisionless Boltzmann equation}
We denote by $f(\vec{q},\vec{k},\tau) \equiv f_0(q) [1+\Psi(q,\vec{k},\mu,\tau)]$ the perturbed NCDM phase-space distribution at conformal time $\tau$, with comoving wavenumber $\vec{k}$ and comoving momentum\footnote{While we keep equations general in our derivation, when implementing our approximation for massive neutrinos, we follow the \textsc{class} convention and work with comoving momenta and energies rescaled by their temperature today.} $\vec{q}$. Throughout we denote by $\epsilon(q,\tau)=\sqrt{q^2+a^2m^2}$ the comoving energy, and define $\mu \equiv \hat{k} \cdot \hat{q}$. In the synchronous gauge, the collisionless Boltzmann equation governing the evolution of the phase-space perturbation $\Psi(q,\vec{k},\mu,\tau)$ is \cite{Ma:1995ey}
\begin{eqnarray}
\dot{\Psi}+ik\mu\frac{q}{\epsilon}\Psi + \frac{d\ln f_0}{d\ln q}\left[ \dot{\eta} - \frac{\dot{h} + 6 \dot{\eta}}{2} \mu^2 \right] = 0,
\label{eq:Boltzmann}
\end{eqnarray}
where overdots denote derivatives with respect to conformal time $\tau$. 
The source terms $\dot{h}(\vec{k},\tau)$ and $\dot{\eta}(\vec{k},\tau)$ are the derivatives of the synchronous gauge metric perturbations with respect to conformal time.

We may decompose $\Psi(\mu)$ on the basis of Legendre polynomials as usual through
\cite{Ma:1995ey}
\be
\Psi(q, \vec{k}, \mu, \tau) = \sum_{\ell} (-i)^\ell (2 \ell +1) \Psi_{\ell}(\vec{k}, q, \tau) P_\ell(\mu). \label{eq:Legendre-exp}
\ee
Only the first three multipole moments are directly relevant for the Einstein field equations.

\subsection{Small-scale quasi-stationary approximation}

While the metric perturbations are themselves sourced by NCDM perturbations through the perturbed Einstein field equations, we may still formally treat them as external sources. Our goal is then as follows: given an ``exact" solution for $\Psi$ up to some time $\tau_*$, and given metric perturbations at all times, we seek an approximate solution at subsequent times $\tau \geq \tau_*$. 

\begin{figure}[ht!]
\includegraphics[width = \linewidth,trim= 0 10 0 10]{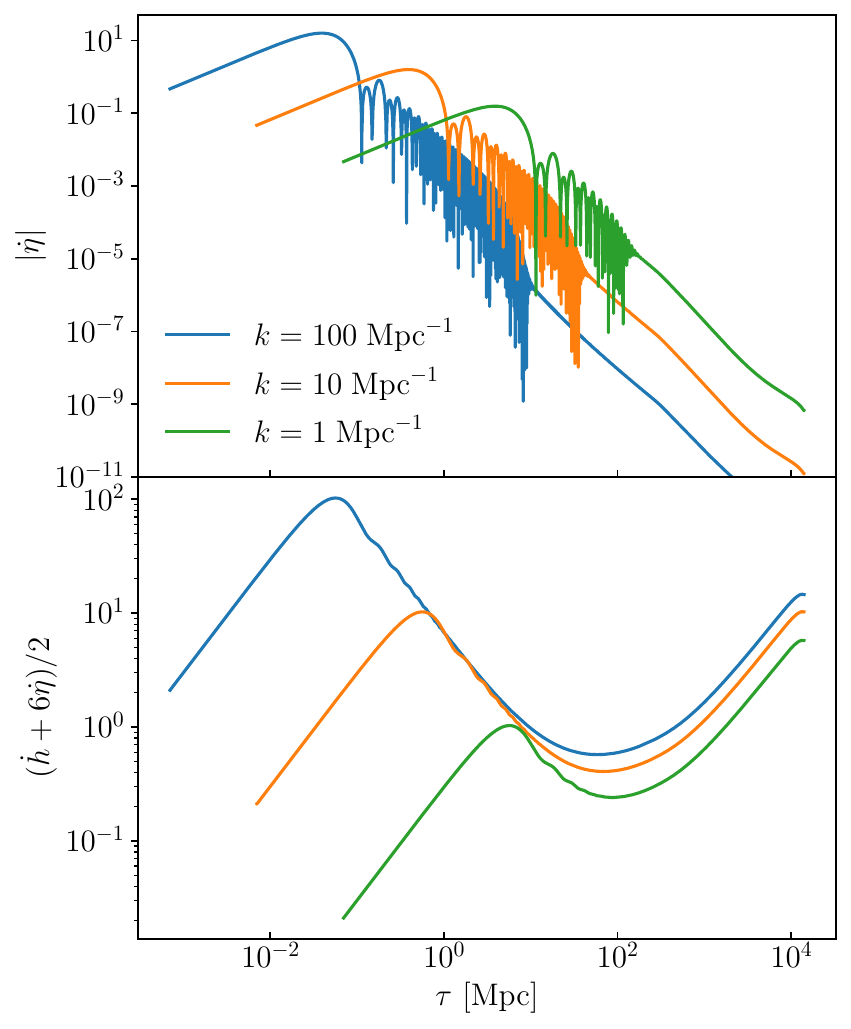}
\caption{Source terms in the Boltzmann equation in the synchronous gauge: $\dot{\eta}$ (top) which decays quickly and $(\dot{h}+6\dot{\eta})/2$ (bottom) which determines the smooth late-time behavior of neutrino perturbations. The relatively small rapid oscillations on top of a smooth behavior in $(\dot{h}+6\dot{\eta})/2$ allow us to find an accurate quasi-stationary analytic approximation for NCDM perturbations.}
\label{fig:hprime-6etaprime}
\end{figure}

Our small-scale approximation relies on two key observations:\\
$\bullet$ After horizon entry, $\dot{\eta}$ is a rapidly oscillating function with a quickly decaying amplitude $\dot{\eta}(k \tau \gg 1) \sim \sin(k \tau/\sqrt{3})/(k \tau)^2$ during radiation domination, and vanishing during matter domination, as we show in Appendix \ref{app:small-scale} (see the top panel of Fig.~\ref{fig:hprime-6etaprime}). At times $\tau \gg 1/k$ we may thus simply neglect the term $\dot{\eta}$ in the Boltzmann equation for $\Psi$.\\
$\bullet$ The combination $\dot{h}+6\dot{\eta}$ is mostly a smooth function of time, varying on a Hubble timescale, as derived in Appendix \ref{app:small-scale} and illustrated in the bottom panel of Fig.~\ref{fig:hprime-6etaprime}. This function also shows rapid oscillations superimposed on top of its running mean, but their amplitude is much smaller than the overall magnitude of $\dot{h} + 6 \dot{\eta}$. Note that the relative amplitude of oscillations is slightly smaller for that combination of metric terms than for $\dot{h}$ alone.

For sufficiently small scales, the source term $\dot{h} + 6 \dot{\eta}$ thus varies on a much longer timescale than the time it takes to free-stream across a comoving wavelength. We may therefore seek a quasi-stationary (QS) particular solution $\Psi_{\rm QS}$ of Eq.~\eqref{eq:Boltzmann}, for which $\partial_\tau \Psi_{\rm QS} \sim a H \Psi_{\rm QS} \ll (k q/\epsilon) \Psi_{\rm QS}$. This quasi-stationary solution is to be summed with a homogeneous solution to smoothly connect with the exact solution at $\tau_*$.

\begin{figure*}[ht!]
\includegraphics[width = 0.329\linewidth,trim= 0 10 0 10]{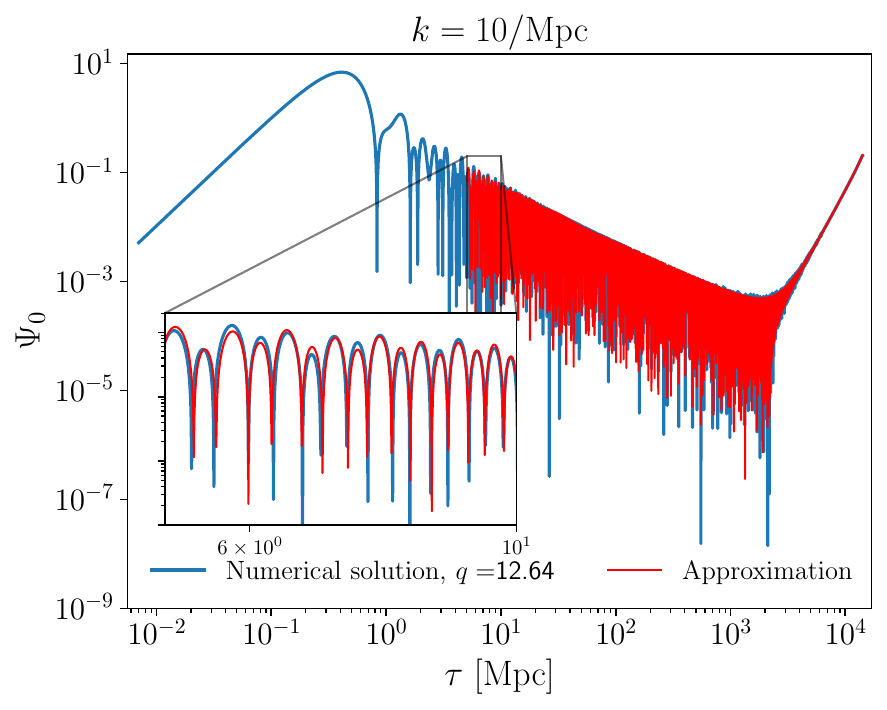}
\includegraphics[width = 0.329\linewidth,trim= 0 10 0 10]{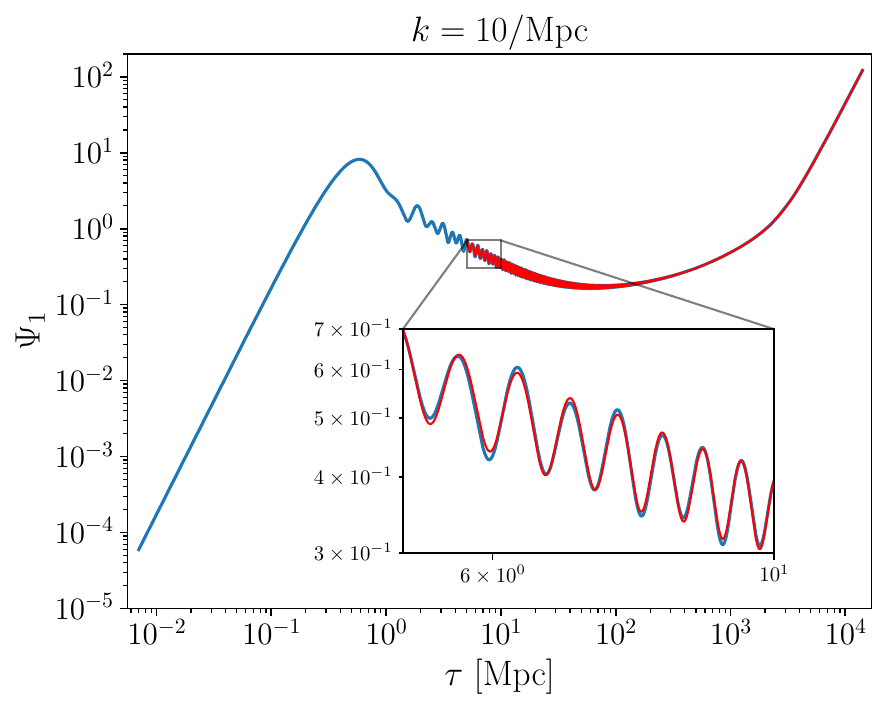}
\includegraphics[width = 0.329\linewidth,trim= 0 10 0 10]{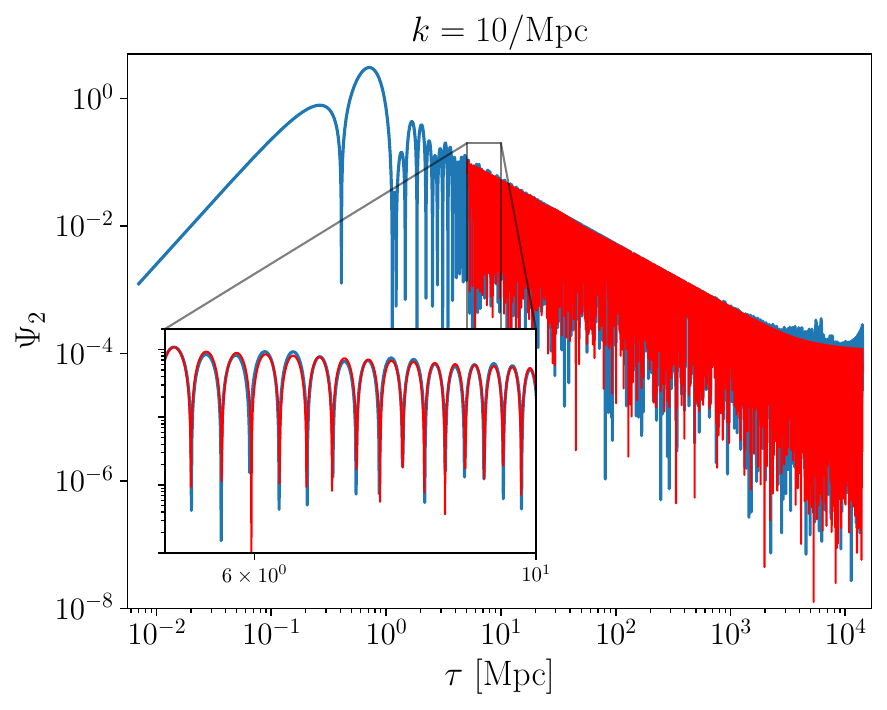}
\caption{The first three multipole moments of massive neutrino phase-space distribution function perturbations, $\Psi_0$, $\Psi_1$, and $\Psi_2$, for comoving wavenumber $k = 10$ Mpc$^{-1}$ and comoving momentum $q = 12.64 ~T_0$, where $T_0$ is the neutrino temperature today. Blue: numerical solutions from \texttt{CLASSIER}; Red: small-scale approximations from Eqs.~\eqref{eq:Psi0-ss}--\eqref{eq:Delta_l}.}
\label{fig:high-k-approx}
\end{figure*}

To derive the solution in a most compact form, for a fixed $(k, q)$, we start by defining the dimensionless variable
\be
\xi \equiv k q \int_{\tau_*}^{\tau} \frac{d\tau'}{\epsilon(q, \tau')}.
\ee
Physically, this variable represents the ratio of the comoving distance traveled by a particle of momentum $q$ between $\tau_*$ and $\tau$ to the comoving lengthscale of the perturbation. We then rewrite the Boltzmann equation for $\Psi$ (neglecting the metric term $\dot{\eta}$) in the form
\be
\Psi'(\xi) + i \mu \Psi = S(\xi) \mu^2, 
\ee
where primes denote derivatives with respect to $\xi$, and
\begin{eqnarray}
S(\xi) &\equiv& \frac{\epsilon}{kq} \frac{d \ln f_0}{d \ln q} \frac{\dot{h} + 6 \dot{\eta}}{2}.
\end{eqnarray}
The slow variation of the source term implies that $S'(\xi)/S(\xi) \sim (aH) \epsilon/kq \equiv s \ll 1$, and we may seek quasi-stationary solutions as a series expansion in that small parameter. We found that it is sufficient to go to second order in $s$, for which the quasi-stationary approximation is 
\barr
\Psi_{\rm QS}(\xi, \mu) = - i S(\xi) \mu + S'(\xi).
\label{eq:sol-QSS}
\earr
A full solution that is continuous at $\tau_*$ (corresponding to $\xi_* = 0$) is then easily obtained by adding the homogeneous solution $e^{- i \mu \xi}$ with the adequate amplitude:
\barr
\Psi^{\rm approx}(\xi, \mu) &\equiv& \Psi_{\rm QS}(\xi, \mu) \nonumber\\
&+& e^{- i \mu \xi}\Big[ \Psi(\xi_*, \mu) - \Psi_{\rm QS}(\xi_*, \mu)\Big].
\earr
This expression can also be obtained starting from the integral solution of the differential equation, integrating by parts twice, and neglecting the remainder proportional to $S''$. This full solution is our complete small-scale approximation. Its explicit first three multipole moments are
\barr
\Psi_0^{\rm approx}(\xi) &=& \Psi_{{\rm QS},0}(\xi) + \sum_{\ell} (-1)^\ell (2 \ell + 1) j_{\ell}(\xi) \Delta_{\ell}(\xi_*), \label{eq:Psi0-ss} \\
\Psi_1^{\rm approx}(\xi) &=& \Psi_{{\rm QS},1}(\xi) - \sum_{\ell} (-1)^{\ell}(2 \ell + 1) j_{\ell}'(\xi) \Delta_{\ell}(\xi_*), ~~~~~~\label{eq:Psi1-ss}\\
\Psi_2^{\rm approx}(\xi) &=& \Psi_{{\rm QS},2}(\xi) \nonumber\\
&& + \sum_{\ell} (-1)^{\ell}(2 \ell + 1) \frac{3 j_{\ell}''(\xi) + j_\ell(\xi)}{2} \Delta_{\ell}(\xi_*),\label{eq:Psi2-ss}
\earr
where 
\barr
\Delta_\ell(\xi_*) &\equiv& \Psi_\ell(\xi_*) - \Psi_{\rm QS,\ell}(\xi_*), \label{eq:Delta_ell}\\
\Psi_{{\rm QS},0}(\xi) &=& S'(\xi),\\
\Psi_{{\rm QS},1}(\xi) &=& \frac 13 S(\xi),\\
\Psi_{{\rm QS},2}(\xi) &=& 0.\label{eq:Delta_l}
\earr
Equations \eqref{eq:Psi0-ss}--\eqref{eq:Delta_l} constitute our main result.\footnote{The explicit forms of the small-scale approximation [Eqs.~\eqref{eq:Psi0-ss}--\eqref{eq:Psi2-ss}] suggest a potential extension: one could directly rewrite the NCDM contributions to the metric source terms in Einstein equations using these expressions, thereby bypassing the explicit evolution of the NCDM multipole hierarchy. Such a strategy could yield further efficiency gains while broadening the applicability of the method to non-standard dark matter scenarios with distinctive small-scale signatures. However, we do not consider the implementation of this idea in \texttt{CLASSIER} as it will require solving different differential equations for different momentum bins making the implementation complex, unless the approximation we develop is valid for all momentum bins.} We checked that including the neglected $\dot{\eta}$ term makes virtually no difference on any observables.

Note that all the terms involving the spherical Bessel functions vanish at $\xi \rightarrow \infty$, so that at late times, the approximation reduces to $\Psi(\xi,\mu) \approx \Psi_{\rm QS}(\xi,\mu)$ as expected. Given that, at late times, $\dot{\eta} \rightarrow 0$, but $\dot{h}$ grows, the dominant asymptotic terms are
\begin{eqnarray}
\Psi_0 &\approx&  \frac12 \frac{d\ln f_0}{d\ln q} \frac{\epsilon~ \partial_{\tau}(\epsilon \dot{h})}{(kq)^2} , \nonumber\\
\Psi_1 &\approx& \frac16 \frac{d\ln f_0}{d\ln q} \frac{\epsilon \dot{h}}{k q},\nonumber\\
\Psi_2 &\rightarrow& 0.
\label{eq:Psi-late}
\end{eqnarray}
Note that the late-time behavior of $\Psi_0$ above matches the analytic result of Ref.~\cite{Nascimento:2023psl} [Eq.~(A22)], in the Newtonian gauge (see also Ref.~\cite{Ringwald:2004np} for the asymptotic expression of density perturbations). For completeness, we provide our small-scale approximation in the Newtonian gauge in Appendix.~\ref{appendix:Newtonian}.

The accuracy of our small-scale approximation in Eqs.~\eqref{eq:Psi0-ss}--\eqref{eq:Delta_l} is demonstrated in Fig.~\ref{fig:high-k-approx}, which compares it with fully numerical solutions for $\Psi_0, \Psi_1, \Psi_2$, shown for comoving wavenumber $k=10\,{\rm Mpc}^{-1}$ and comoving momentum $q\simeq 12~ T_0$, where $T_0$ is the neutrino temperature today. The blue curves are full numerical solutions obtained using \texttt{CLASSIER}, which were shown to agree with high-precision \texttt{CLASS} calculations \cite{Lee:2025zym}, while the red curves show our small-scale approximations. The figure illustrates that our method accurately reproduces both the late-time behavior of massive neutrino perturbations and their oscillatory features arising from the homogeneous solution encoding early-time contributions. Although the oscillating part near the switch at $\tau_*$ shows slight differences with the numerical solution, these small differences in rapidly oscillating terms have very little impact on the matter power spectrum, as we show explicitly in the next section. 

Note that the small-scale approximation can even surpass the accuracy of the fully numerical solution at late times (see the right panel of Fig.~\ref{fig:high-k-approx}). This occurs because the accuracy of numerical solutions inevitably depends on numerical settings [e.g., in the case of \texttt{CLASSIER}, parameters controlling the non-uniform FFT (NUFFT) resolution], whereas the analytic small-scale approximation converges to the true solution in the regime where its validity conditions are well satisfied.

The small-scale approximation we develop here can easily be applied to massless species (e.g., massless neutrinos). In the massless limit, we have $\epsilon=q$ and $\xi=k(\tau-\tau_*)$. We provide the small-scale approximation in this limit, in Appendix.~\ref{appendix:massless}

\section{Implementation and results}
\label{sec:implementation-results}

\subsection{Implementation in \texttt{CLASSIER}}

We implement the small-scale approximation developed in Section~\ref{sec:approximation} into \texttt{CLASSIER}, a code that replaces the traditional truncated Boltzmann hierarchy for NCDM perturbations with a set of integral equations evaluated iteratively for self-consistency. In this work, we apply the approximation to the standard case of massive neutrinos.

In \texttt{CLASS}, the parameter \texttt{ppr->ncdm\_fluid\allowbreak\_trigger\allowbreak\_tau\allowbreak\_over\allowbreak\_tau\_k} (default value 31) sets the conformal time at which the fluid approximation for NCDM begins, i.e.\ $\tau = 31/k$. The same prescription applies in the zeroth iteration of \texttt{CLASSIER}. For improved accuracy, we instead adopt $\tau=51/k$ and increase the truncation order of the Boltzmann hierarchy to $\ell_{\rm max}^{\rm NCDM}=60$ (default 17 in \texttt{CLASS}). These modifications add negligible cost but improve the accuracy of the subsequent integral-equation iterations. In practice, the zeroth iteration evolves the truncated hierarchy up to $\tau=51/k$, after which the calculation switches to the fluid approximation. Later iterations evaluate the integral solutions for neutrino perturbations as convolutions using \texttt{FINUFFT}
(Flatiron Institute Nonuniform Fast Fourier Transform) \cite{Barnett:2019qbm,Barnett:2020nufft_aliasing}\footnote{\href{https://finufft.readthedocs.io/en/latest/}{https://finufft.readthedocs.io/en/latest/}}. 

As explained earlier, our small-scale approximation is valid as long as the parameter $s \equiv a H \epsilon/kq$ is small, i.e.~as long as the perturbation comoving wavelength is small relative to the comoving distance free-streamed in a Hubble time. While the free-streaming particles are relativistic, this parameter is simply $s \approx aH/k$, and decreases monotonically with time. However, once particles become non-relativistic, $s \approx a^2 H m/(k q)$, which increases during matter domination. Therefore, it is possible for some momentum bins $q$ to satisfy $s \ll 1$ for some period of time, but no longer once the particles become sufficiently non-relativistic.

We could in principle switch off the small-scale approximation at a different time for each momentum $q$ and wavenumber $k$. However, this would significantly complicate our implementation, with little gains in terms of overall computational efficiency. Therefore, we instead choose to use the small-scale approximation for a given $(q, k)$ pair only if it holds at the present time $\tau_0$, i.e.~only if $s \ll 1$ at $\tau_0$. If this is indeed the case, we switch on the small-scale approximation at $\tau_* = 51/k$, and use it until the present time.

\begin{figure}[ht!]
\includegraphics[width = \linewidth,trim= 0 10 0 10]{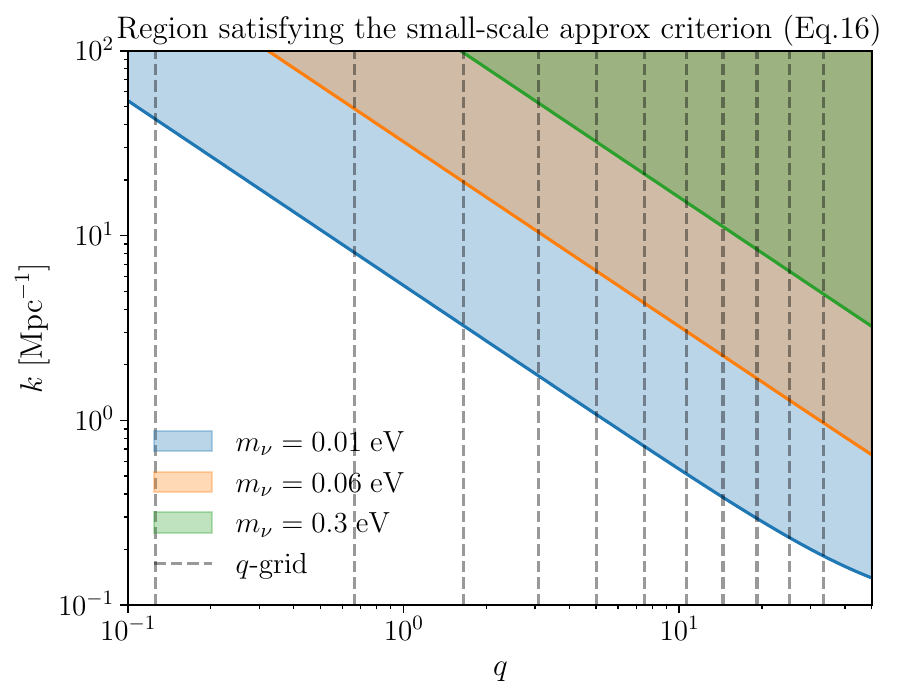}
\caption{Regions in $(k,q)$ space that satisfy the small-scale approximation criterion of Eq.~\eqref{eq:criterion}. Different colors correspond to different neutrino masses. Vertical dashed lines indicate the $q$-grid used. Here $q$ is the comoving momentum in units of $T_0$, the neutrino temperature today.}
\label{fig:criterion}
\end{figure}

Concretely, we found that the small-scale approximation yields better than 1\% accuracy in $\Psi_0$ today provided that the following criterion is satisfied:
\begin{equation}
s(k, q, \tau_0) \equiv \frac{H_0 \epsilon(q, \tau_0)}{kq} < 0.0025.
\label{eq:criterion}
\end{equation}
We show in Fig.~\ref{fig:criterion} the regions of $(k, q)$ parameter space where the criterion \eqref{eq:criterion} holds.
Whenever the criterion \eqref{eq:criterion} is satisfied, we replace the full integral evaluation with the small-scale approximation, Eqs.~\eqref{eq:Psi0-ss}--\eqref{eq:Psi2-ss}, where the initial multipoles $\Psi_\ell(\tau_*)$ are evaluated at $\tau_*\equiv51/k$ for $\ell \leq \ell_{\rm max}^{\rm NCDM}$.

The small-scale approximation is straightforward to compute but involves repeated evaluations of spherical Bessel functions $j_\ell(\xi)$, of their first and second derivatives. Since these functions only need to be computed once, we precompute them on a grid of 5000 equally spaced values of $\xi$ over $0 \leq \xi < 10^3$ and store them in external file (\texttt{ncdmfft/j\_ell.txt}). When the small-scale approximation is enabled, the code loads these tables and interpolates as needed, eliminating the overhead of repeatedly evaluating the kernel functions during runtime.

To further reduce the computational cost without loss of accuracy, at $\xi > 1000$, we switch from the full small-scale approximation [Eqs.~\eqref{eq:Psi0-ss}--\eqref{eq:Delta_l}] to these simplified late-time expressions [Eq.~\eqref{eq:Psi-late}]. 

We note that in the original \texttt{CLASSIER} implementation the NUFFT grids were chosen conservatively to resolve the rapid oscillations induced by large momentum bins. With the small-scale approximation handling precisely these bins, such high grid resolution is no longer necessary. Accordingly, when the approximation is active the grid density is adaptively reduced for large-$k$ modes, providing an additional runtime saving on top of the analytic speed-up.\footnote{We also introduce a new parameter \texttt{k\_ncdmfft\_min}, set to $0.5\,{\rm Mpc}^{-1}$ in this work. For $k$ below this threshold, the code reverts to the standard \texttt{CLASS} treatment with the Runge–Kutta integrator (\texttt{evolver=0}). Under the current precision setting, the standard \texttt{CLASS} method can appear faster than \texttt{CLASSIER}, partly because of the relatively small number of momentum bins and partly due to the present structure of the convolution step in \texttt{CLASSIER}. This latter effect mainly impacts low-$k$ modes, and could potentially be mitigated by restructuring the code in future revisions.}

Finally, note that in principle, one could iterate on the procedure we described above, i.e.~start with standard \texttt{CLASS}, with the fluid approximation, to compute the metric perturbations, use \texttt{CLASSIER} with the small-scale approximation to compute NCDM perturbations, store the latter, and re-compute metric perturbations sourced by these more accurate NCDM perturbations, etc... In practice, we have found that one single iteration is enough to reach sub-0.1\% accuracy on the matter power spectrum, as we describe further in the next section.

\begin{figure*}[ht!]
\includegraphics[width = \linewidth,trim= 0 10 0 10]{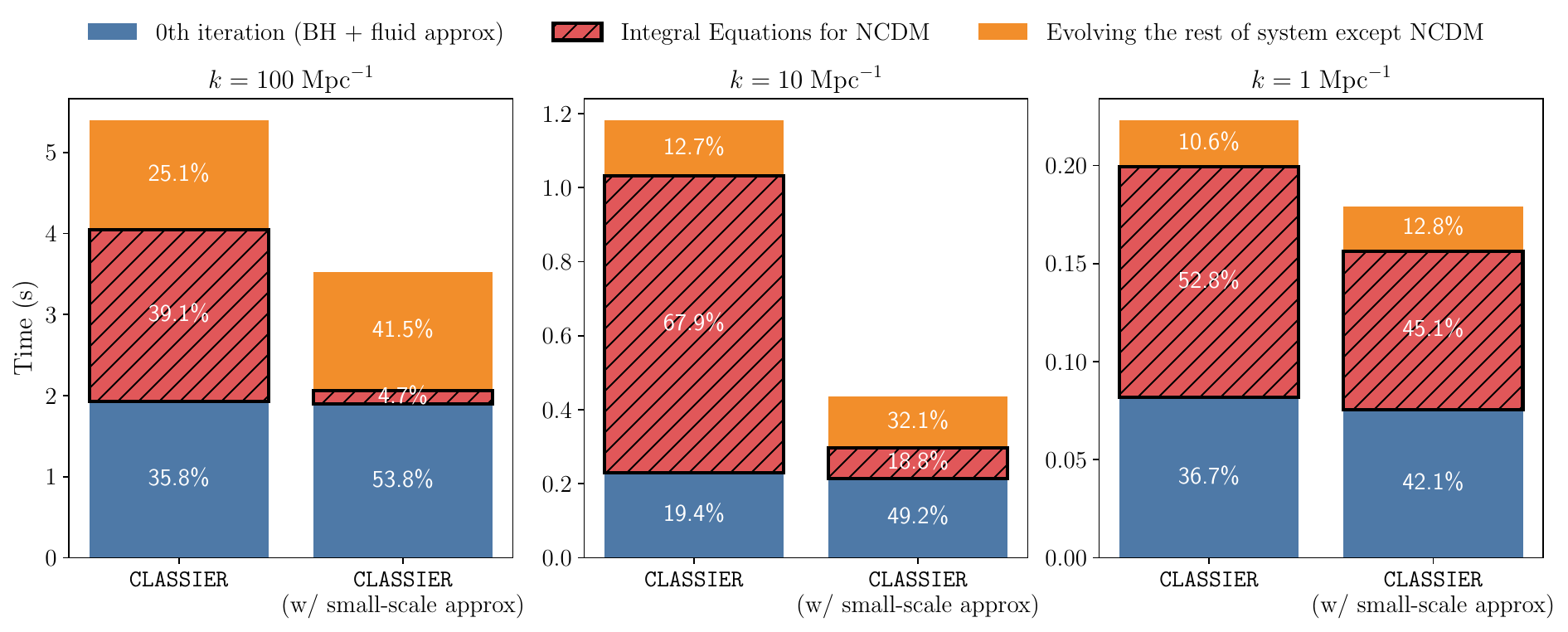}
\caption{Runtime distribution for evolving perturbations of three representative $k$-modes in \texttt{CLASSIER} (left: $k=100\;{\rm Mpc}^{-1}$, middle: $k=10\;{\rm Mpc}^{-1}$, right: $k=1\;{\rm Mpc}^{-1}$), shown without (left bar) and with (right bar) the small-scale approximation. Each bar is split into contributions from the zeroth-iteration step that solves the truncated Boltzmann hierarchy (BH) with the late-time fluid approximation (blue), from the integral-equation evaluation of NCDM perturbations (red hatched), and from the rest of the system (orange). The small-scale approximation substantially reduces the time spent on the NCDM integral-equation step, which otherwise accounts for a non-negligible share of the runtime at high $k$. At small $k$, fewer momenta qualify, but these modes contribute negligibly to the overall runtime. When the approximation is used, we reduce the resolution of $\tau$-grid for convolution calculations and this additionally saves the small amount of runtime.}
\label{fig:runtime-dist}
\end{figure*}

\subsection{Runtime improvements}

We now assess the performance of the small-scale approximation developed in this work, both in terms of computational efficiency and accuracy of cosmological observables. Throughout the runtime and accuracy comparisons, we adopt the default cosmological parameters of \texttt{CLASS v3.2.2} with one massive neutrino of mass $0.06\,\mathrm{eV}$, and use the \texttt{CLASS}:UHP precision setting of Ref.~\cite{Euclid:2024imf}, unless stated otherwise.

Figure~\ref{fig:runtime-dist} shows the runtime distribution for three representative $k$-modes ($100, 10$, and $1\;{\rm Mpc}^{-1}$), separated into contributions from the zeroth-iteration step that solves the truncated Boltzmann hierarchy for NCDM with the late-time fluid approximation (blue), from the integral-equation evaluation (red hatched), and from the rest of the system (orange). The small-scale approximation substantially reduces the time required for the integral-equation step, which otherwise accounts for a noticeable fraction of the runtime at high $k$. At lower $k$, fewer momentum bins satisfy the criterion of Eq.~\eqref{eq:criterion} (see Fig.~\ref{fig:criterion}), so the runtime reduction is less pronounced; however, these low-$k$ modes contribute little to the overall runtime budget. Taken together, these results indicate that the small-scale approximation delivers savings close to the maximum possible in the portion of the calculation dominated by NCDM perturbations.

Table~\ref{tab:runtime} summarizes the runtime performance for different choices of $k_{\rm max}$, for \texttt{CLASS}, the original version of \texttt{CLASSIER}, and with the small-scale approximation described in the present work. For completeness, results under a lower reference precision setting, corresponding to the \texttt{CLASS} default precision with two modifications, \texttt{l\_max\allowbreak\_ncdm=40} ($\ell_{\rm max}^{\rm NCDM}=40$) and \texttt{ncdm\allowbreak\_fluid\allowbreak\_approximation=3} (no fluid approximation), are given in Appendix~\ref{appendix:runtime}.

\begin{table}[!ht]
  \centering
  \begin{tabular}{c|c|c|c}
  $k_{\rm max}$ & \texttt{CLASS} & \texttt{CLASSIER} & \texttt{CLASSIER} \\
  $({\rm Mpc}^{-1})$ & ~~$\ell_{\rm max}^{\rm NCDM}=40$~~ & (original) & w/ small-scale approx.\\
  \hline\hline
  $10$  & 16~sec  & 12~sec  & \textbf{5~sec} \\
  \hline
  $50$  & 80~sec  & 30~sec   & \textbf{15~sec} \\
  \hline
  $100$ & 160~sec & 49~sec  & \textbf{26~sec} \\
  \end{tabular}
  \caption{Runtimes (four threads, 16-inch MacBook Pro--M4 chip, Nov 2024 release) under the \texttt{CLASS}:UHP precision setting, for different $k_{\rm max}$. The  small-scale approximation leads to a reduction in total runtime relative to the original version of \texttt{CLASSIER} ranging from $\sim$ 45\% to 60\%. With the small-scale approximation, \texttt{CLASSIER} is faster than \texttt{CLASS} by a factor of 3--6. Single-thread runtimes are roughly four times longer.} \label{tab:runtime}
\end{table}

These results show that the small-scale approximation substantially reduces the cost of evaluating NCDM perturbations in \texttt{CLASSIER}. For example, for $k_{\rm max}=100\,{\rm Mpc}^{-1}$ the total runtime decreases by $46\%$ relative to the baseline \texttt{CLASSIER} runs. Similar improvements appear with lower $k_{\rm max}$, with reductions of order $50\%$ across the board. Importantly, these gains are achieved while preserving the accuracy of the solutions (see below). The breakdown in Fig.~\ref{fig:runtime-dist} clarifies that the overall savings are driven primarily by reductions in the NCDM integral-equation step, which constitutes a noticeable fraction of the runtime at high $k$. We also have checked that applying the small-scale approximation to massless neutrinos as shown in Appendix.~\ref{appendix:massless} can further reduce the total runtime by up to 20\%.\footnote{This feature is not included in the current public release of \texttt{CLASSIER}. The reason is that small numerical artifacts arising during the tight-coupling approximation in \texttt{CLASS}—when both $\ell_{\rm max}^{\rm NCDM}$ and $\ell_{\rm max}^{\rm ur}$ (the truncation scale of the momentum-integrated Boltzmann hierarchy for massless neutrinos) are increased—become amplified within the iterative integral-equation scheme. We plan to include this refinement in a future release once this issue is resolved.}

Throughout this work, \texttt{CLASSIER} employs the big-$q$ approximation of Ref.~\cite{Lee:2025zym},\footnote{At early times, for sufficiently large comoving momenta $q$, the perturbations remain effectively relativistic so their $q$-dependence factors out of the source functions. The convolution integrals are then evaluated only for the largest $q$-bin and the results for other bins are obtained by rescaling, avoiding redundant calculations.} which already lowers the cost of NCDM evaluations. The runtime reductions reported here are therefore an additional improvement on top of that baseline.

\subsection{Accuracy of the matter power spectrum}

\begin{figure}[ht!]
\includegraphics[width = \linewidth,trim= 0 10 0 10]{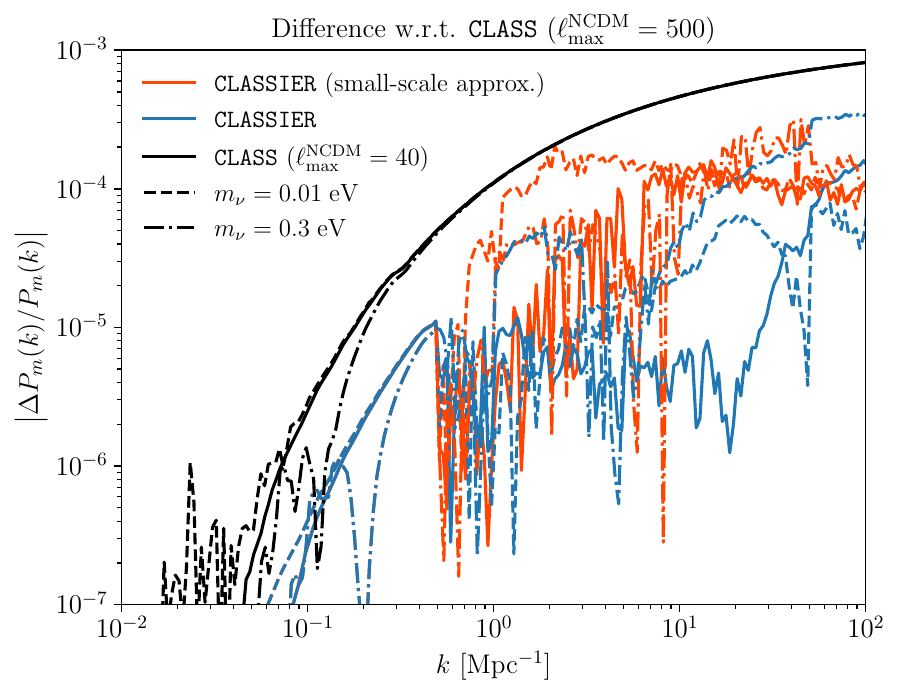}
\caption{Absolute fractional differences of the $z=0$ matter power spectrum with respect to \texttt{CLASS} with $\ell^{\rm NCDM}_{\rm max}=500$. The solid lines are with the neutrino mass $m_\nu=0.06$eV, and the dashed and the dot-dashed lines are with $m_\nu=0.01$eV and $0.3$eV, respectively.}
\label{fig:DPm}
\end{figure}

To validate accuracy, we compute the $z=0$ matter power spectrum and compare it against a high-precision \texttt{CLASS} calculation with $\ell_{\rm max}^{\rm NCDM}=500$. Figure~\ref{fig:DPm} shows the absolute fractional differences. For reference, we also show standard \texttt{CLASS} with $\ell_{\rm max}^{\rm NCDM}=40$, as well as \texttt{CLASSIER} without the small-scale approximation. This figure demonstrates that our small-scale approximation (orange curve) maintains accuracy at the sub-0.1\% level across the full $k$-range considered, performing essentially indistinguishably from the original integral solution (blue curve). Thus, the small-scale approximation reproduces the correct late-time behavior and the characteristic free-streaming suppression of power, leaving cosmological observables unaffected. We repeat this test with two additional neutrino mass values, $0.01$ eV and $0.3$ eV, and present with dashed and dot-dashed lines, respectively. This demonstrates that the accuracy of the small-scale approximation remains the same for a wide range of neutrino mass.

In summary, the small-scale approximation roughly halves the runtime of \texttt{CLASSIER} while preserving the accuracy of the matter power spectrum, making it well-suited for applications that require repeated evaluations of cosmological perturbations, such as parameter inference in upcoming surveys.

\section{Conclusion}
\label{sec:conclusion}

We have derived a highly accurate analytic approximation for small-scale non-cold relic (NCDM) perturbations using quasi-stationary particular solutions to the collisionless Boltzmann equation and implemented it in \texttt{CLASSIER}. Combined with the homogeneous solution which carries the early-time contribution, the approximation replaces costly late-time convolutions with compact analytic expressions. The approximation accurately captures the smooth late-time behavior of NCDM perturbations, in addition to the rapid oscillatory features due to free-streaming.

On the performance side, the approximation reduces the total runtime in \texttt{CLASSIER} by roughly a factor of two across a broad range of $k_{\rm max}$ (Table~\ref{tab:runtime}). Compared to standard \texttt{CLASS} runs (with $\ell_{\rm max}^{\rm NCDM}=40$ and no late-time massive neutrino fluid approximation) under the same precision setting, \texttt{CLASSIER} with this approximation is faster by a factor of 3--6. On the accuracy side, comparisons to high-precision \texttt{CLASS} calculations show sub-$0.1\%$ agreement in the matter power spectrum today (Fig.~\ref{fig:DPm}), faithfully capturing the small-scale suppression from free-streaming. The method therefore achieves substantial speed-ups with no degradation in accuracy, making it particularly well suited for parameter inference pipelines in the precision cosmology era.

The validity of the approximation is governed by a simple scale-dependent condition in Eq.~\eqref{eq:criterion} expressed in terms of $s \equiv aH\epsilon/kq$, which quantifies the ratio of the perturbation wavelength to the distance free-streamed per Hubble time. Even if a species becomes non-relativistic, the quasi-stationary solution remains accurate on sufficiently small scales provided $s$ is small enough throughout the evolution. Within this regime, the late-time behavior of the leading multipoles is controlled by simple $h$-derivative terms [Eq.~\eqref{eq:Psi-late}].

Although we have benchmarked the numerics for the canonical case of massive neutrino, the approach applies to exotic non-cold relics \cite{DEramo:2020gpr,Banerjee:2025gwe,Venumadhav:2015pla,Abazajian:2019ejt,Bansal:2024afn,Alvey:2021sji} and with some modifications to models with dark-matter decays to non-cold relics \cite{Aoyama:2014tga,FrancoAbellan:2021sxk,Bencke:2025}.  The improved precision and efficiency in the wavenumber range $1\lesssim (k/h\,{\rm Mpc}^{-1}) \lesssim 100$ may be useful for recent models \cite{Sobotka:2024tat,Co:2025lrd}.

We release an implementation in \texttt{CLASSIER} that precomputes the required kernels, adapts grids when the approximation is active, and falls back seamlessly to the full integral evaluation when the criterion is not met. We expect the approximation to deliver increasing benefits as upcoming analyses push to smaller scales and as likelihood pipelines demand ever larger numbers of forward model evaluations.

\section*{Acknowledgements}

We thank J.~Lesgourgues, C.~Nascimento, and Y.~Wong for useful discussions. This work was supported at JHU by NSF Grant No.\ 2412361, NASA ATP Grant No.\ 80NSSC24K1226, and the Templeton Foundation. NL was supported by the Horizon Fellowship from Johns Hopkins University.

\section*{Data Availability}

The data that support the findings of this article are openly available \cite{CLASSIER}.

\appendix

\section{Small-scale behavior of synchronous-gauge potentials} \label{app:small-scale}

In this appendix we derive the small-scale behavior of synchronous-gauge potentials $\dot{\eta}$ and $\dot{h}$. We do so starting from known properties of the Newtonian-gauge potentials $\phi$, $\psi$ and change gauge to obtain $\dot{\eta}$, $\dot{h}$. The gauge-transformation equations for the metric perturbations $\eta$ and $\alpha \equiv \dot{h} + 6 \dot{\eta}$ are \cite{Ma:1995ey}
\barr
\psi = \frac1{2 k^2} \frac1{a} \frac{d}{d\tau}(a \alpha),\ \ \ \ \phi = \eta - \frac{a H}{2 k^2} \alpha. \label{eq:gauge-trans}
\earr

$\bullet$ In the radiation era, neglecting anisotropic stress and neutrino contributions, it is well known that 
\be
\phi(k, \tau) \approx \psi(k, \tau) \approx 3 \phi_i(k) \frac{j_1(x)}{x}, \ \ \ x \equiv k \tau/\sqrt{3}. \label{eq:phi-approx}
\ee
Using the fact that $a \propto \tau \propto x$ in the radiation era, we may solve Eqs.~\eqref{eq:gauge-trans} explicitly and obtain
\barr
\alpha &\approx& 6 \sqrt{3} k \phi_i(k) \frac{ 1 - j_0(x)}{x}, \\ 
\eta &\approx& 3 \phi_i(k) \left[\frac{1 - j_0(x)}{x^2} + \frac{j_1(x)}{x} \right], 
\earr
implying
\be
\dot{\eta} \approx \sqrt{3} k \phi_i(k) \left[2\frac{j_0(x) -1 - x j_1(x)}{x^3} + \frac{j_0(x)}x \right].
\ee
We see that, at late times $k \tau \gg 1$ (but still in the radiation era), $\dot{\eta} \propto \sin(k \tau/\sqrt{3})/(k \tau)^2$. We also see that the oscillatory part of $\alpha$ is suppressed relative to its running mean by a factor of order $\sim 1/(k \tau) \ll 1$ at late times.

$\bullet$ During matter domination, $\phi \approx \psi \approx \textrm{constant}$ and $a \propto \tau^2$, implying $\alpha \approx (\phi/3) \tau$ and $\eta \approx \phi/3$, hence $\dot{\eta} \approx 0$.

\newpage

\section{Small-scale approximation\\in the Newtonian Gauge}
\label{appendix:Newtonian}

\begin{figure*}[ht!]
\includegraphics[width = 0.329\linewidth,trim= 0 10 0 10]{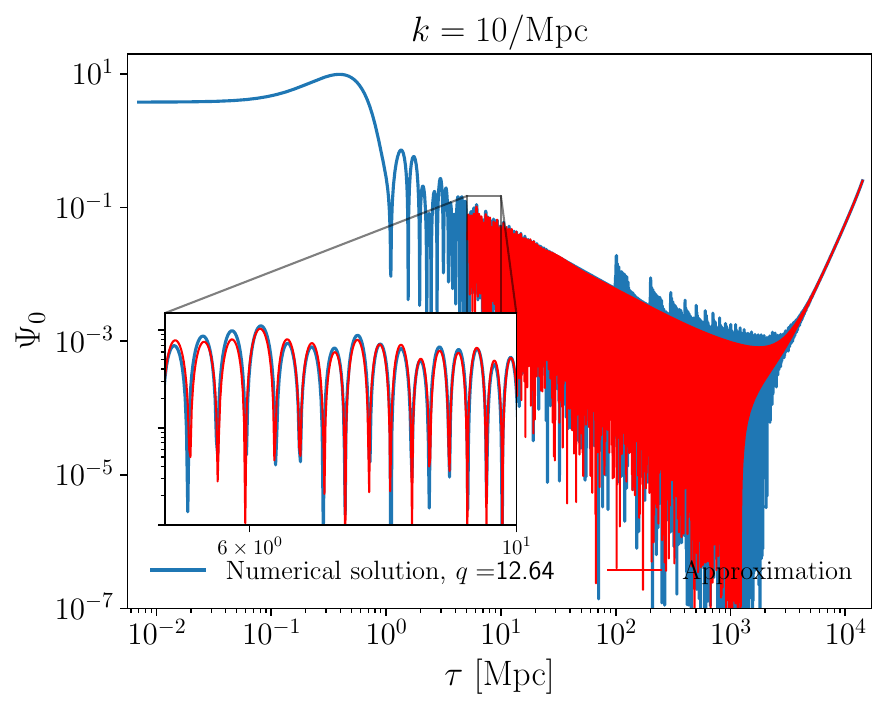}
\includegraphics[width = 0.329\linewidth,trim= 0 10 0 10]{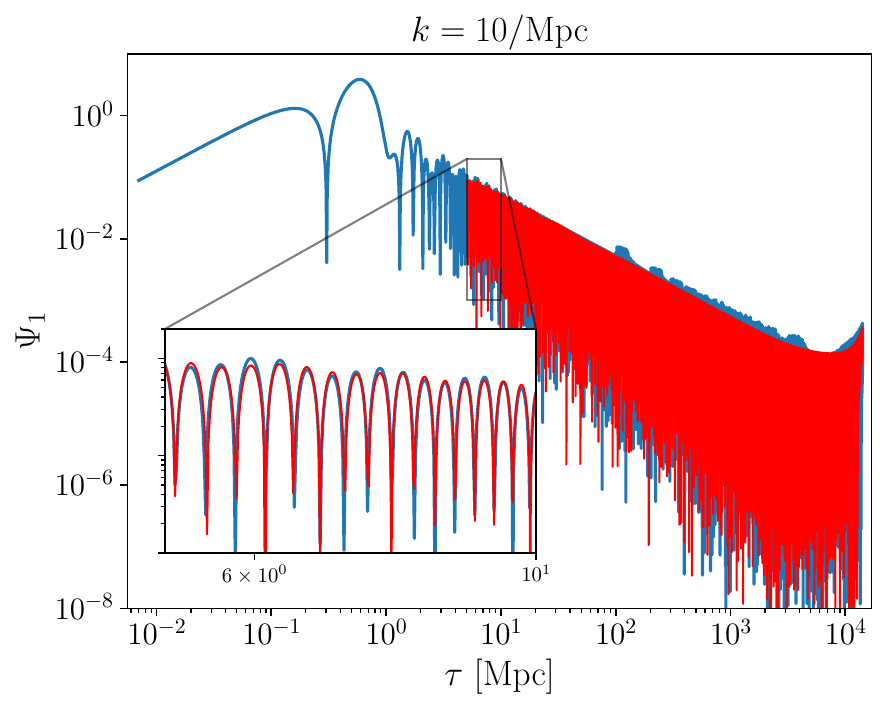}
\includegraphics[width = 0.329\linewidth,trim= 0 10 0 10]{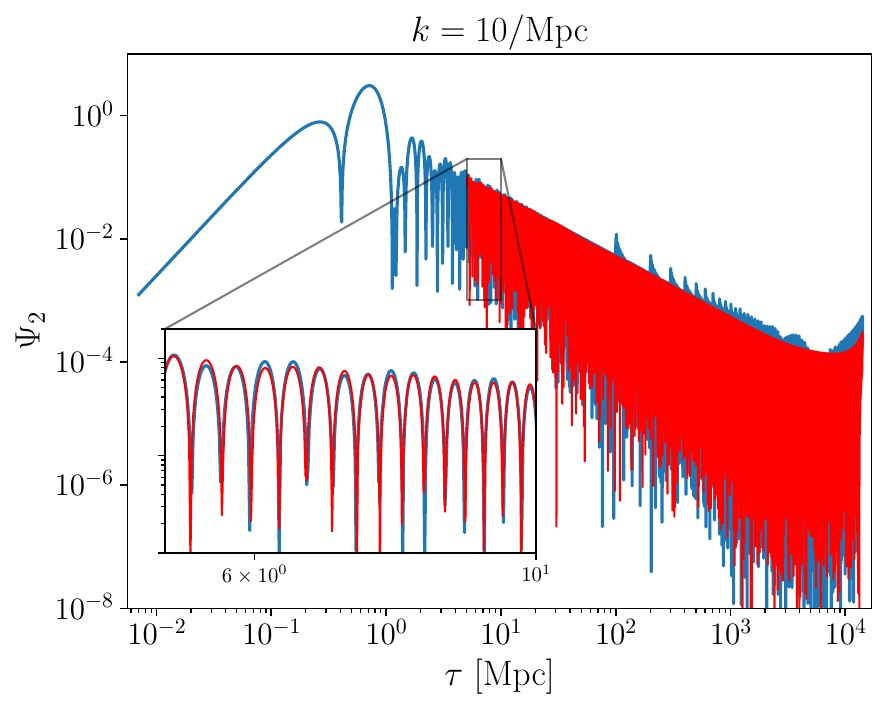}
\caption{Similar plots to Fig.~\ref{fig:high-k-approx}, but shown in the Newtonian gauge. Blue: numerical solutions from \texttt{CLASS} with $\ell_{\rm max}^{\rm NCDM}=500$ and no fluid approximation for massive neutrinos; Red: small-scale approximations from Eqs.~\eqref{eq:Psi0-ss}--\eqref{eq:Delta_l} and \eqref{eq:Delta_l-Newtonian}. The recurring features appearing at $\tau \sim 2\ell_{\rm max}^{\rm NCDM}/k \simeq 10,{\rm Mpc}$ in the blue curves are known numerical artifacts arising from the truncation of the Boltzmann hierarchy \cite{Ma:1995ey,Lesgourgues:2011rh}. These artifacts are absent in Fig.~\ref{fig:high-k-approx}, where the numerical solutions were obtained using \texttt{CLASSIER}, which is free from such truncation-induced errors \cite{Lee:2025zym}.}
\label{fig:high-k-approx-Newtonian}
\end{figure*}

In this appendix, we provide the small-scale approximation in the Newtonian gauge. The collisionless Boltzmann equation in the Newtonian gauge is \cite{Ma:1995ey}
\begin{eqnarray}
\dot{\Psi}+ik\mu\frac{q}{\epsilon}\Psi + \frac{d\ln f_0}{d\ln q}\left[ \dot{\phi} - i\frac{\epsilon k}{q} \psi \mu \right] = 0.
\end{eqnarray}
If we take the same approach in this gauge, we encounter ill-behaved quasi-stationary solutions inversely proportional to $\mu$. To regularize this singularity, we introduce a collisional damping term on the right-hand-side of the Boltzmann equation, proportional to a small coefficient $0 < \alpha \ll 1$ and think of the collisionless equation as the limit of this collisional system when $\alpha \rightarrow 0^+$.\footnote{This method to regularize singularities is similar in spirit to that taken to compute resonant torques in a disk around a central mass orbited by a satellite \cite{1987Icar...69..157M}.} For a fixed $k$ and $q$, we rewrite this equation in terms of the dimensionless variable $\xi$. We are thus looking to solve the following differential equation for $\Psi(\xi)$: 
\be
\Psi'(\xi) + i( \mu -i\alpha)\Psi =  iS(\xi)\mu, 
\ee
where primes denote derivatives with respect to $\xi$, and
\begin{eqnarray}
S(\xi) &\equiv& \left(\frac{\epsilon}{q}\right)^2 \frac{d \ln f_0}{d\ln q}\psi.
\end{eqnarray}
We checked that the contribution from the $\dot{\phi}$ term is negligible, suppressed by $(aH/k)\ll1$ on subhorizon scales relative to that from the $\psi$ term. At second order in this small parameter $s=aH\epsilon/kq$, we obtain the quasi-stationary solution as
\begin{equation}
\Psi_{\rm QS}(\xi, \mu) \approx S(\xi) + i\frac{S'(\xi) }{\mu-i\alpha},~~~~\label{eq:sol-QSS-Newtonian}
\end{equation}
where we have only kept the small parameter $\alpha$ in the terms where it is explicitly needed to regularize the divergence at $\mu \rightarrow 0$. 

Just like for the synchronous gauge described in the main text, our approximation consists in adding to the QS solution a homogeneous solution with the adequate amplitude to be continuous at $\tau_*$. Computing the $\ell = 0, 1, 2$ multipole moments and taking the limit $\alpha \rightarrow 0^+$, our approximation takes the form of Eqs.~\eqref{eq:Psi0-ss}-\eqref{eq:Delta_ell}, with
\barr
\Psi_{{\rm QS},0}(\xi) &=&  S(\xi)-\frac{\pi}{2} S'(\xi),\nonumber\\
\Psi_{{\rm QS},1}(\xi) &=&  -S'(\xi),\nonumber\\
\Psi_{{\rm QS},2}(\xi) &=& -\frac{\pi}{4}S'(\xi).\label{eq:Delta_l-Newtonian}
 \earr
The approximations are compared with full numerical solutions in Fig.~\ref{fig:high-k-approx-Newtonian}. Note that this comparison is performed by a standalone python script, and the approximations haven't been implemented in \texttt{CLASSIER}. We plan to implement these approximations in the Newtonian gauge into \texttt{CLASSIER}, once the integral-equation approach developed in Ref.~\cite{Lee:2025zym} becomes available in that gauge in the near future.

\section{Small-scale approximation\\in massless limit}
\label{appendix:massless}

In the massless limit $q/\epsilon = 1$ and the perturbation to the phase-space density only depends on momentum through the prefactor $d \ln f_0/d \ln q$:
\be
\Psi_\ell(\vec{k}, q, \mu, \tau) = \frac{d \ln f_0}{d \ln q} ~ \widetilde{\Psi}(\vec{k}, \mu, \tau). \label{eq:Psi-tilde}
\ee

Following Ref.~\cite{Ma:1995ey}, we may integrate out the $q$-dependence in the phase-space distribution perturbation:
\begin{eqnarray}
F (\vec{k},\mu,\tau) \equiv \frac{\int q^2 dq \;q f_0(q) \Psi(\vec{k},q,\mu,\tau)}{\int q^2 dq \;q f_0(q)} = - 4 \widetilde{\Psi}(\vec{k}, \tau),~~
\end{eqnarray}
where in the second equality we inserted Eq.~\eqref{eq:Psi-tilde} and integrated by parts, assuming $f_0(q \rightarrow \infty) = 0$. The small-scale approximation for the first three multipoles $F_\ell(\vec{k}, \tau)$ then takes the same form as Eqs.~\eqref{eq:Psi0-ss}--\eqref{eq:Delta_l}, with $\xi = k(\tau - \tau_*)$, $\Psi_\ell(\xi_*) \rightarrow F_\ell(\xi_*)$, and $S \rightarrow - 2(\dot{h} + 6 \dot{\eta})/k$.

\section{Runtime under lower reference precision setting}
\label{appendix:runtime}

For completeness, Table~\ref{tab:runtime-default} shows runtime comparisons under a lower reference precision setting, based on the \texttt{CLASS} default precision but with two modifications: \texttt{l\_max\_ncdm=40} ($\ell_{\rm max}^{\rm NCDM}=40$) and \texttt{ncdm\allowbreak\_fluid\allowbreak\_approximation=3} (no fluid approximation).\\

\begin{table}[!ht]
  \centering
  \begin{tabular}{c|c|c|c}
  $k_{\rm max}$ & \texttt{CLASS} & \texttt{CLASSIER} & \texttt{CLASSIER} \\
  $({\rm Mpc}^{-1})$ & ~~$\ell_{\rm max}^{\rm NCDM}=40$~~ & (original) & w/ small-scale approx.\\
  \hline\hline
  $10$  & 2.6~sec  & 1.9~sec  & \textbf{1.2~sec} \\
  \hline
  $50$  & 11~sec  & 3.9~sec   & \textbf{2.2~sec} \\
  \hline
  $100$ & 21~sec & 6.0~sec  & \textbf{3.4~sec} \\
  \end{tabular}
  \caption{Runtimes (four threads, 16-inch MacBook Pro--M4 chip, Nov 2024 release) under a lower reference precision setting, corresponding to the \texttt{CLASS} default precision setting except \texttt{l\_max\_ncdm=40} ($\ell_{\rm max}^{\rm NCDM}=40$) and \texttt{ncdm\allowbreak\_fluid\allowbreak\_approximation=3} (no fluid approximation), for different $k_{\rm max}$. The  small-scale approximation leads to a reduction in total runtime relative to the original version of \texttt{CLASSIER}  $\sim$ 40\%. With the small-scale approximation, \texttt{CLASSIER} is faster than \texttt{CLASS} by a factor of 2--6. Single-thread runtimes are roughly four times longer.} \label{tab:runtime-default}
\end{table}

\pagebreak

\bibliography{mybib}

\end{document}